\documentclass[comment, prl,twocolumn,nofootinbib, preprintnumbers, superscriptaddress]{revtex4}

\usepackage{amsmath,amssymb,bm,slashed,braket}
\usepackage{graphicx}
\usepackage{epstopdf}
\usepackage{float}
\usepackage[colorlinks=true,
            linkcolor=blue,
            urlcolor=blue,
            citecolor=green,          
            bookmarks=true,
            bookmarksnumbered=true,
            breaklinks=true,
            pdfpagemode=Fullscreen,
            pdfstartview=FitBH]{hyperref}

\usepackage[normalem]{ulem}
\usepackage{color}

\definecolor{Orange}{cmyk}{0,0.61,0.87,0}
\definecolor{JungleGreen}{cmyk}{0.99,0,0.52,0}
\definecolor{OliveGreen}{cmyk}{0.64,0,0.95,0.40}
\definecolor{Brown}{cmyk}{0,0.81,1,0.60}
\definecolor{RoyalBlue}{cmyk}{0.71,0.53,0,0.12}
\definecolor{Gray}{cmyk}{0,0,0,0.40}
\definecolor{LightPink}{cmyk}{0.0,0.25,0,0}
\definecolor{LLightPink}{cmyk}{0.0,0.10,0,0}
\definecolor{LightBlue}{cmyk}{0.25,0,0,0}
\definecolor{LightGray}{cmyk}{0,0,0,0.2}

\usepackage{xcolor}
\definecolor{gesfpurple}{rgb}{0.47,0.19,0.42}

\definecolor{gesflanse}{rgb}{0.00,0.50,0.50}

\definecolor{gesfblue}{rgb}{0.08,0.42,0.76}

\definecolor{gesfred}{rgb}{1,0,0}

\definecolor{gesfwhite}{rgb}{1,1,1}

\definecolor{gesfblack}{rgb}{0,0,0}

\newcommand{\geqn}[1]{Eq.\,\hypersetup{linkcolor=blue}(\ref{#1})\hypersetup{linkcolor=blue}}
\newcommand{\gfig}[1]{{\hypersetup{linkcolor=violet}Fig.\,\ref{#1}\hypersetup{linkcolor=blue}}}

\graphicspath{{figs/}}

\begin{document}

\title{
Dark Matter Annihilation via Breit-Wigner Enhancement with Heavier Mediator
}

\author{Yu Cheng}
\email[Corresponding Author: ]{chengyu@sjtu.edu.cn}
\affiliation{Tsung-Dao Lee Institute \& School of Physics and Astronomy, Shanghai Jiao Tong University, China}
\affiliation{Key Laboratory for Particle Astrophysics and Cosmology (MOE) \& Shanghai Key Laboratory for Particle Physics and Cosmology, Shanghai Jiao Tong University, Shanghai 200240, China}

\author{Shao-Feng Ge}
\email{gesf@sjtu.edu.cn}
\affiliation{Tsung-Dao Lee Institute \& School of Physics and Astronomy, Shanghai Jiao Tong University, China}
\affiliation{Key Laboratory for Particle Astrophysics and Cosmology (MOE) \& Shanghai Key Laboratory for Particle Physics and Cosmology, Shanghai Jiao Tong University, Shanghai 200240, China}

\author{Jie Sheng}
\email[Corresponding Author: ]{shengjie04@sjtu.edu.cn}
\affiliation{Tsung-Dao Lee Institute \& School of Physics and Astronomy, Shanghai Jiao Tong University, China}
\affiliation{Key Laboratory for Particle Astrophysics and Cosmology (MOE) \& Shanghai Key Laboratory for Particle Physics and Cosmology, Shanghai Jiao Tong University, Shanghai 200240, China}

\author{Tsutomu T. Yanagida}
\email{tsutomu.yanagida@sjtu.edu.cn}
\affiliation{Tsung-Dao Lee Institute \& School of Physics and Astronomy, Shanghai Jiao Tong University, China}
\affiliation{Key Laboratory for Particle Astrophysics and Cosmology (MOE) \& Shanghai Key Laboratory for Particle Physics and Cosmology, Shanghai Jiao Tong University, Shanghai 200240, China}

\begin{abstract}

We propose a new scenario that both the dark matter
freeze-out in the early Universe and its possible annihilation 
for indirect detection around a supermassive black hole 
are enhanced by a Breit-Wigner resonance. With the mediator mass 
larger than the total initial dark matter mass, this annihilation is almost forbidden at late times. Thus,
the stringent cosmic microwave background and indirect detection constraints
do not apply. However, a 
supermassive black hole can accelerate the dark matter
particles to reactivate this resonant annihilation whose subsequent decay to 
photons leaves a unique signal. 
The running Fermi-LAT and the future COSI satellites can test this scenario.

\end{abstract}

\maketitle 

{\bf Introduction} --
More than 80\% of the matter in our Universe today 
is dark matter (DM) \cite{Young:2016ala,Arbey:2021gdg}.
One major hunting strategy is the DM direct detection
that uses the DM scattering with nucleus or electron.
Currently, the Xenon-based experiments
have reached ton-scale
\cite{PandaX-4T:2021bab,XENON:2022ltv,LZ:2022lsv}.
However, for the non-relativistic DM in our galaxy only those with 
mass $\gtrsim \mathcal{O}(1)$\,GeV
carry large enough kinetic energy to
make the nuclear recoil from elastic scattering to 
overcome the detection threshold.
While the DM above GeV scale is already highly 
constrained,
the sub-GeV region still has large parameter space \cite{Schumann:2019eaa}.
Thus, light DM is becoming more popular nowadays \cite{Cooley:2022ufh}.

In the standard freeze-out scenario, 
the thermally averaged DM annihilation
cross section $\langle \sigma v \rangle$ for obtaining 
the observed relic abundance should be around
$\langle \sigma v \rangle \sim 10^{-26}\,$cm$^{3}$s$^{-1}$
\cite{Steigman:2012nb}. The same annihilation into the 
Standard Model (SM) particles can still 
happen at late times.
Its subsequent electromagnetic energy injection into the environment modifies 
the ionization history of the Universe and finally affects the 
observed cosmic microwave background (CMB).
Thus, the DM freeze-out scenario receives stringent
constraint from CMB
\cite{Hansen:2003yj,Bean:2003kd,Pierpaoli:2003rz,Chen:2003gz,Padmanabhan:2005es,Slatyer:2009yq,Steigman:2015hda,Roszkowski:2017nbc,Planck:2018nkj,Cang:2020exa,Kawasaki:2021etm,Liu:2023nct}.

One solution is  
the forbidden-type DM \cite{Griest:1990kh,DAgnolo:2015ujb,Delgado:2016umt,DAgnolo:2020mpt,Wojcik:2021xki}
whose annihilation is kinematically prohibited 
at late times.
Unfortunately, this also makes it difficult 
to leave indirect detection signals today.
The only place for the forbidden annihilation
to re-open is around a supermassive black hole (SMBH) \cite{Cheng:2022esn,Cheng:2023hzw}
since the strong gravitational force
can accelerate the DM particles to overcome the annihilation threshold. 
The $p$-wave annihilation can also be enhanced around SMBH
\cite{Amin:2007ir,Shelton:2015aqa,Arina:2015zoa,Johnson:2019hsm,Chiang:2019zjj}.

In this paper, we propose an alternative scenario
based on the Breit-Wigner resonance with a heavy mediator
to escape the CMB constraint
and leave indirect detection signals around the SMBH.
Instead of a generic discussion on the $s$-channel resonance \cite{Griest:1990kh,Liu:2017lpo}, the original Breit-Wigner mechanism using a light mediator (the mediator mass $m_\phi$ is smaller than $2 m_{\chi}$
where $m_\chi$ is the DM mass) was invented
to enhance the annihilation at late times to explain the 
electron-positron excess observed by the
PAMELA \cite{PAMELA:2008gwm}, ATIC \cite{Chang:2008aa},
PPB-BETS \cite{PPB-BETS:2008zzu}, and AMS-02 \cite{AMS:2013fma}
cosmic-ray observations.
This excess can also be explained by pulsars
\cite{Hooper:2008kg,Profumo:2008ms,Malyshev:2009tw,Yuan:2013eja,Yin:2013vaa,Gaskins:2016cha}.
We do the opposite with a heavy mediator, $m_\phi > 2 m_{\chi}$.
The DM annihilation
during freeze-out is enhanced by 
its thermal energy that 
can compensate the mass difference 
to reach 
the $s$-channel resonance pole.
When temperature cools down, the 
DM annihilation moves away from 
the resonance pole and becomes 
greatly suppressed at late times 
to escape the CMB constraint \cite{Bai:2012qy}.

This new Breit-Wigner scenario
with a heavy mediator can 
reactivate around an SMBH and
the subsequent decay of the 
final-state SM particles can 
leave a unique signature. 
In addition to the existing gamma-ray telescopes, such as
Fermi-LAT \cite{Fermi-LAT:2009ihh} and H.E.S.S. \cite{HESS:2018pbp}
that can search for DM signals around the SMBH Sgr\,$A^*$ in the energy range from $\mathcal{O}(100)\,$MeV to TeV,
the upcoming telescope Compton Spectrometer and Image (COSI)
aims at detecting the soft gamma-ray of $0.2 \sim 5\,$MeV \cite{Tomsick:2021wed}. 
This opens a new window for testing our new scenario.

{\bf DM Production with Breit-Wigner Resonance} --
In the Breit-Wigner mechanism, the DM $\chi$ with mass
$m_\chi$ annihilates into SM particles
through an $s$-channel mediator $\phi$.
Although the mediator can have arbitrary spins
in principle, we assume a scalar mediator 
for simplicity. The general form of the annihilation cross 
section with a Breit-Wigner resonance is \cite{Ibe:2008ye,Guo:2009aj},
\begin{eqnarray}
  \sigma
=
  \frac{16 \pi \omega \beta_f}{s \bar{\beta}_i \bar{\beta}_f  \beta_i} 
  \frac{m_\phi^2 \Gamma^2_{\phi}}{\left(s-m_\phi^2\right)^2+m_\phi^2 \Gamma^2_{\phi}} B_i B_f.
\label{eq:crosssection1}
\end{eqnarray}
The factor $\omega \equiv P_\chi (2 J_\phi + 1)/(2 J_\chi +1)^2$ is determined
by the spins of the mediator ($J_\phi$) and
the initial-state DM ($J_\chi$), respectively,
as well as the
symmetry factor $P_\chi$  for 
the mediator decay into a pair of identical DM
particles ($\phi \rightarrow \chi \chi$).
With a scalar mediator and a Majorana DM implemented
in this paper, $\omega = 1/2$.
The initial/final-state
phase space factor
$\beta_{i(f)} \equiv \sqrt{1 - 4 m_{i(f)}^2/ E_{\rm cm}^2}$
is evaluated with the 
center-of-mass energy $ \sqrt{s}
\equiv E_{\rm cm}$ and 
$\bar \beta_{i(f)} \equiv \sqrt{1 - 4 m_{i(f)}^2/ m_\phi^2}$
with the mediator mass $m_\phi$.
Here, $m_{i(f)}$ is the mass of initial (final)
particles.
Around the resonance,
these two factors coincides with each other,
$\beta_{i(f)} \approx \bar \beta_{i(f)}$.
The branching ratios 
$B_{i(f)} \equiv \Gamma_{i(f)}/\Gamma_{\phi}$ are defined for the mediator $\phi$ decay into the initial-
and final-state particles with decay width $\Gamma_{i}$
and $\Gamma_f$, respectively, with
the total decay width
$\Gamma_\phi$.
A large boost factor for
the indirect detection signal can be achieved with a light mediator,
$m_\phi < 2 m_{\chi}$,
in the original Breit-Wigner scenario \cite{Ibe:2008ye,Guo:2009aj}. Since $s$ is always 
larger than $m_\phi^2$, it keeps
decreasing and approaching 
the pole when temperature cools
down as shown in the Fig.1 of \cite{Ibe:2008ye}. 

\begin{figure}
    \centering
    \includegraphics[width=0.486
    \textwidth]{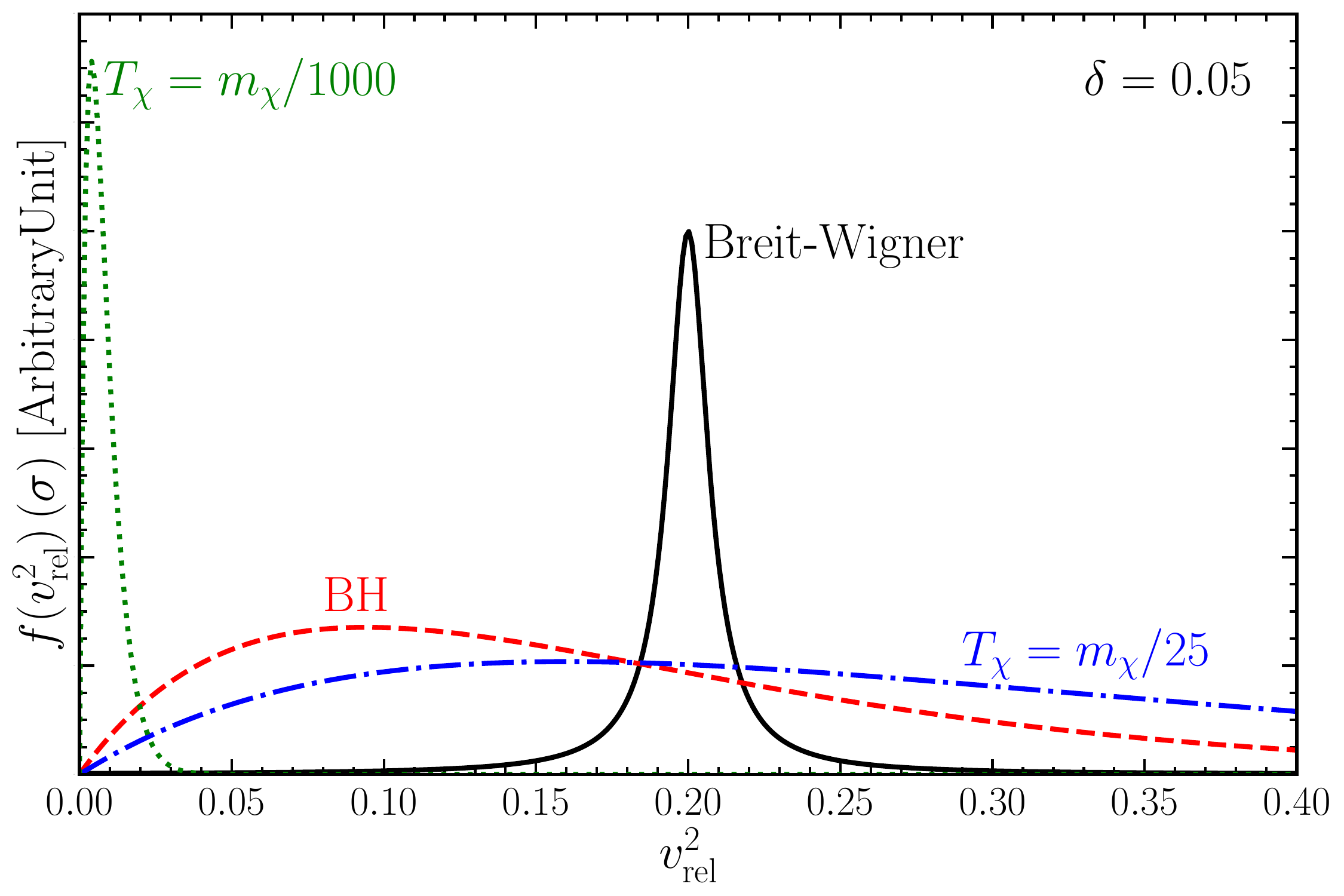}
    \caption{
    The overlap between different DM phase space 
    distribution functions $f (v_{\rm rel}^2)$ and the DM annihilation
    cross section $\sigma$ as a function of the relative velocity
    squared
    $v^2_{\rm rel}$.
    The black curve clearly shows
    the Breit-Wigner resonance with a 
    positive mass difference 
    $\delta = 0.05$. The blue dash-dotted
    and green dotted curves are the 
    DM velocity distributions 
    when the DM temperature is just 4\% ($x = 25$ which is equivalently $T_\chi = m_\chi / 25$) or 0.1\% ($x =1000$ or $T_\chi = m_\chi / 1000$) of 
    its mass around freeze-out or at the late Universe, respectively.
    The red dashed curve
    shows the case around an SMBH at the place
    with the largest DM annihilation rate per unit radius. }
    \label{fig:BWCurve}
\end{figure}

Our scenario takes instead a heavy mediator, 
$m_\phi > 2 m_\chi$, for the resonance.
In this case, the $s$-channel resonance can be
achieved only when the DM has 
high enough kinetic energy.  
To make the resonance feature
transparent, we parametrize 
the mediator mass as
$  m_\phi^2
\equiv
  4 m^2_\chi (1 + \delta)$
with $\delta$ denoting 
the mass difference between $m_\phi$ and
$2 m_\chi$ while the non-relativistic center-of-mass energy squared
$s \approx 4 m_\chi^2+m_\chi^2\left(\vec{v}_{\rm rel}\right)^2$ is a function of the relative velocity $\vec{v}_{\rm rel } \equiv \vec{v}_1-\vec{v}_2$. 
Then the cross section in
\geqn{eq:crosssection1} becomes, 
\begin{equation}
  \sigma
=
  \frac{16 \pi \omega \beta_f}{s \bar{\beta}_i \bar \beta_f \beta_i}
  \frac{\gamma_\phi^2}{\left( 
  \frac{\vec{v}_{\mathrm{rel}}^2/4- \delta}{1 + \delta} \right)^2 + \gamma_\phi^2} B_i B_f,
\label{eq:sigma}
\end{equation}
with the mediator decay width
$\Gamma_\phi$ has been normalized as
$\gamma_\phi \equiv \Gamma_{\phi}/m_\phi$. 
Since this normalized mediator decay width $\gamma_\phi$ is proportional to the square of the coupling between the mediator $\phi$ and DM $\chi$ as well as SM particles, its magnitude can equivalently represent the magnitude of the cross section shown in \geqn{eq:sigma}.
For a heavy mediator, $\delta > 0$, the cross section reaches the 
Breit-Wigner resonance pole when $|\vec v_{\rm rel}|^2/4 =  \delta$.
For illustration, the black curve in \gfig{fig:BWCurve} shows that
the annihilation cross section with $\delta = 0.05$ peaks at exactly 
$v_{\rm rel}^2 =  0.2$ as expected. 

In the early Universe, 
the DM number density evolution is determined by the thermally 
averaged annihilation cross
section,
\begin{equation}
    \langle \sigma v_{\rm rel} \rangle
\equiv
     \frac{x^{3 / 2}}{2 \pi^{1 / 2}} \int_0^{\infty} d v_{\mathrm{rel}} v_{\mathrm{rel}}^2\left(\sigma v_{\mathrm{rel}}\right) e^{-x v_{\mathrm{rel}}^2 / 4},
\label{eq:ThermalAverageNonrel}
\end{equation}
where $x \equiv m_\chi / T_\chi$ parametrizes the DM temperature $T_\chi$ and
the DM 
phase space distribution has been approximated by the Maxwell distribution. With 
Breit-Wigner resonance, the thermally averaged $\langle \sigma v_{\rm rel} \rangle$ is maximized 
when the Maxwell distribution (blue dash-dotted)
peak overlaps with the pole (black solid), 
$T_\chi  \simeq \delta \times m_\chi$.
This happens around $x \sim 20$ in the early Universe
with the $\delta = 0.05$ adopted in \gfig{fig:BWCurve}.

With
further decreasing temperature, the velocity 
distribution softens. 
For illustration, the
DM with temperature 
$T_\chi = m_\chi/1000$ (green dotted) has 
almost no overlap with 
the Breit-Wigner resonance (black solid).
At the time of CMB formation, the Universe temperature has already dropped to $\mathcal{O}(1)\,$eV. 
The DM
kinetic energy is then not enough
to activate the Breit-Wigner enhancement with a heavy mediator.
Thus, its annihilation is highly 
suppressed and can naturally escape
the CMB constraint.

A broader Breit-Wigner peak
can increase its overlap with
the low-temperature DM distribution. 
However, the overlap is still small since
the normalized mediator decay width $\gamma_\phi$ 
(equivalently the cross section)
can not be too large 
in order to 
produce the correct relic density.

In the freeze-out scenario, the DM yield, 
$Y \equiv n_\chi/s$ where $n_\chi$ is the DM number density
and $s$ the entropy density, evolves according to
the Boltzmann equation \cite{Kolb:1990vq},
\begin{equation}
  \frac{d Y}{d x}
=
-\frac{\tilde{\lambda}}{x^2}
\left(Y^2-Y_{\mathrm{eq}}^2\right),
\label{eq:BoltzmanEq}
\end{equation}
with
$Y_{\rm eq}
\equiv
  0.145 (g /g_{s *}) x^{3 / 2} e^{-x}$.
The thermally averaged cross section has been redefined as, 
\begin{eqnarray}
\tilde{\lambda} 
\equiv
\sqrt{\frac{\pi}{45}} \frac{g_{s *}}{\sqrt{g_{ *}}}\left[1+\frac{T}{3} \frac{\mathrm{d}}{\mathrm{d} T} \ln \left(g_{s *}\right)\right] m_\chi M_{\rm pl}
\left\langle \sigma v_{\rm rel} \right\rangle,
\end{eqnarray}
where $M_{\rm pl}=1.22 \times 10^{19}$\,GeV
is the Planck mass and 
$g_{*}$ ($g_{\mathrm{s} *}$) the
effective relativistic energy (entropy) degrees of freedom \cite{Hindmarsh:2005ix,Drees:2015exa,Laine:2015kra}.
When 
DM is in thermal equilibrium, its 
yield $Y$ tracks $Y_{\rm eq}$. 
As temperature cools down, the DM number density $n_\chi$ drops
exponentially and it begins to freeze out
once the annihilation rate
$n_\chi \braket{\sigma v_{\rm rel}}$ is comparable 
to the Hubble rate $H$, $n_\chi \braket{\sigma v_{\rm rel}}
\simeq H$. We take the freeze-out 
criterion $Y - Y_{\rm eq} \simeq Y_{\rm eq}$ to estimate the freeze-out point $x_f$ \cite{Kolb:1990vq}, 
\begin{equation}
  x_f
\equiv
  \ln \frac{0.038 g_\chi M_{\rm pl} m_\chi \langle\sigma v_{\rm rel} \rangle}{g_*^{1 / 2} x_f^{1 / 2}},
\end{equation}
where the DM  degrees of freedom $g_\chi = 2$
for a Majorana fermion.
This iterative equation gives that 
the DM freezes out at $x_f \sim 25$. 
By solving the differential equation 
$dY / dx = -\tilde \lambda Y^2/ x^2$ from the freeze-out
point $x_f$ to infinity, we can get the DM yield today, 
\begin{equation}
    Y_\chi = \frac{Y(x_f)}{1 + Y(x_f) \int^{\infty}_{x_f} \frac{\tilde \lambda}{x^2} dx}.
\end{equation}
For non-relativistic DM, its 
relic density $\rho_{\chi} = m_\chi s_0 Y_\chi$ is given by its mass
$m_\chi$ and the entropy density today $s_0 = 2891.2$\,cm$^{-3}$ \cite{ParticleDataGroup:2022pth}.
The cosmological observations indicate
that DM contributes $27\%$
of the critical density today
$\rho_c \approx 0.5 \times 10^{-5} $\,GeV/cm$^3$\cite{Planck:2018vyg,ParticleDataGroup:2022pth}.

\begin{figure}[!t]
\centering
 \includegraphics[width=0.486
 \textwidth]{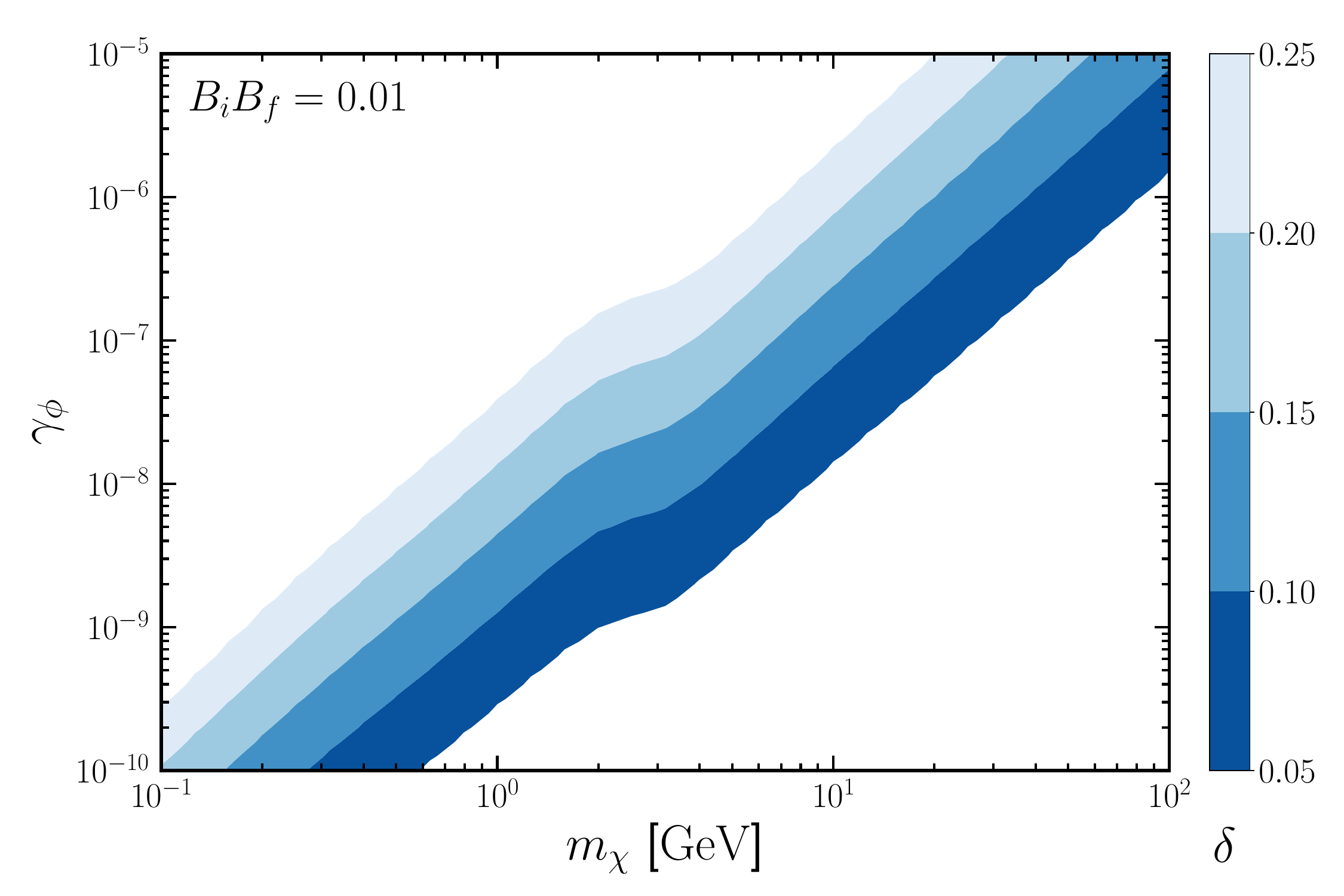}
\caption{
The parameter space to generate the correct 
relic density (27\% of the critical density today) for the
Breit-Wigner resonance scenario
with a heavy mediator.
We fix $B_i B_f = 0.01$ while 
varying the DM mass $m_\chi$ (horizontal 
axis) and the normalized mediator decay width $\gamma_\phi$
(vertical axis). Then, the mass difference $\delta$
is uniquely determined by the DM relic density whose value
can be read off according to the blue color bar
on the right-hand side.
}
\label{fig:RelicDensiy}
\end{figure}

There are four independent parameters, $m_\chi$, $\delta$, $\gamma_\phi$, 
and $B_i B_f$. However, these parameters
are not totally free since the inequality 
$\Gamma^2_{\phi} = (\Gamma_i + \Gamma_f)^2 \geq 4 \Gamma_i \Gamma_f$ requires $B_i B_f \leq 0.25$.
The parameter values
to produce the correct relic
density are shown in 
\gfig{fig:RelicDensiy}. 
For illustration, we fix the 
product of branching ratios $B_i B_f = 0.01$
and vary the DM mass $m_\chi$
as well as the normalized mediator decay width $\gamma_\phi$.
Then, the observed DM relic
density uniquely determines the mass difference $\delta$.
We can first see that 
a larger normalized 
mediator decay width $\gamma_\phi$
and hence a larger DM annihilation cross section
requires a larger DM mass which is the same feature as in the standard case. 
Second, 
a larger mass difference $\delta$ prefers a larger 
normalized mediator decay width $\gamma_\phi$.
This is because a big mass difference to put the
resonance at larger $v^2_{\rm rel}$
makes the DM difficult to touch the Breit-Wigner resonance with a 
heavy mediator. Such difficulty 
can be compensated by a larger coupling strength to make the peak wider with a larger
mediator decay width $\gamma_\phi$.

{\bf The Right-Handed Neutrino DM Model} --
One interesting model to further illustrate our
scenario is a right-handed neutrino (RHN) 
DM \cite{Kusenko:2010ik,Basak:2013cga,Okada:2016tci,Cox:2017rgn,Borah:2018smz,Arcadi:2020aot}. 
The heavy right-handed Majorana neutrinos are widely
considered as a key ingredient beyond the SM. They explain not only the
observed tiny neutrino masses via the seesaw mechanism
\cite{Minkowski:1977sc,Yanagida:1979as,Yanagida:1979gs,Gell-Mann:1979vob}, but also the baryon asymmetry of our Universe
through leptogenesis \cite{Fukugita:1986hr}. Normally, we assume three
heavy RHNs. However, two heavy
Majorana neutrinos are already sufficient to 
explain the baryon asymmetry in our Universe
and the nonzero neutrino mass squared
differences interpreted from oscillation  
\cite{Frampton:2002qc,Raidal:2002xf,Glashow:2003nk,Barger:2003gt,Ge:2010js}.
The remaining right-handed neutrino $N$ can then
serve as a DM candidate. A byproduct is that we have an
anthropic argument \cite{Weinberg:1987dv}
for the presence of three families of quarks and leptons \cite{Ibe:2016yfo}.

We assume a $Z_2$ parity acting only on this
RHN DM $N_\chi$ to make it stable \cite{Cox:2017rgn}.
To ensure thermal equilibrium for $N_\chi$
in the early Universe, 
we introduce a scalar mediator
$\phi$ that couples to $N_\chi$ through a Majorana-type Yukawa term $y\phi N_\chi^T \epsilon N_\chi$ with Yukawa coupling $y$. Notice here that $N_\chi$ is a right-handed two-component
Weyl fermion. 
Since the RHN DM $N_\chi$ is assumed to carry odd parity under $Z_2$,  
$\phi$ is even and can couple with a pair of the SM Higgs bosons, $H$ and $H^{\dagger}$, via $\phi H^{\dagger} H$ with coupling strength $\lambda m_\phi$.
Thus, the interaction Lagrangian is, 
\begin{equation}
  \mathcal L_{\rm int}
=
  ( y \phi N_\chi^T \epsilon N_\chi + h.c.)
+ \lambda m_{\phi} \phi H^{\dagger} H.
\label{Interaction}
\end{equation}

With a heavy mediator, 
$m_\phi > 2 m_\chi$, 
the mediator decay width $\Gamma_\phi$ receives two contributions,
$\Gamma_{\phi} \equiv \Gamma_{\phi \rightarrow N_\chi N_\chi} + \Gamma_{\phi \rightarrow f \bar f}$
for its decay into a pair of DM
($\Gamma_{\phi \rightarrow N_\chi N_\chi}$)
or SM ($\Gamma_{\phi \rightarrow f \bar f}$) fermions, 
\begin{subequations}
\begin{align}
    \Gamma_{\phi \rightarrow N_\chi N_\chi} 
    &=
    y^2 \frac{m_\phi}{4 \pi} 
    \left(1 - \frac{4 m_\chi^2}{m_\phi^2}
    \right)^{3/2},\\
    \Gamma_{\phi \rightarrow f \bar{f}}
    &=
    (\lambda^{\prime})^2
    \frac{m_\phi}{8 \pi} \left(1 - \frac{4 m_f^2}{m_\phi^2}
    \right)^{3/2}.
\end{align}
\end{subequations}
The Yukawa coupling 
between the mediator $\phi$ and the SM fermion $f$,
$\lambda^{\prime}
\equiv m_f \sin \theta /v$, 
is 
a function of the SM fermion mass $m_f$, the Higgs
vacuum expectation value $v$, as well as
the mixing angle $\theta$ between $\phi$ and the SM Higgs particle $h$.
In the limit of heavy mediator,
$m_\phi \gg m_h$, the mixing angle
$\theta 
\equiv
\frac12 \arctan [
\lambda m_\phi v / (m_\phi^2 - m_h^2)]$
reduces to $\theta \approx \lambda v / 2 m_\phi$ and
consequently $\lambda' \approx \lambda m_f / 2 m_\phi$ scales
with the mass ratio of SM fermion and Higgs particle.
The mediator decay branching ratios originally
defined below \geqn{eq:crosssection1} become, 
$B_i \equiv \Gamma_{\phi \rightarrow N_\chi N_\chi} /\Gamma_\phi$
and
$B_f \equiv \Gamma_{\phi \rightarrow f \bar{f}} / \Gamma_\phi$
with $\Gamma_i \equiv \Gamma_{\phi \rightarrow N_\chi N_\chi} $
and $\Gamma_f \equiv \Gamma_{\phi \rightarrow f \bar{f}}$.
The annihilation cross section for
$N_\chi N_\chi \rightarrow f \bar{f}$ can be written as,
\begin{equation}
    \sigma_{N_\chi N_\chi \rightarrow f \bar{f}}
=
  \frac{8 \pi s \beta_i \beta_f^3 }{ m_\phi^4 \bar{\beta}_i^3 \bar{\beta}_f^3  } 
  \frac{m_\phi^2 \Gamma^2_{\phi}}{\left(s-m_\phi^2\right)^2+m_\phi^2 \Gamma^2_{\phi}} B_i B_f.
  \label{eq:crosssection2}
\end{equation}
This formula is consistent with \geqn{eq:crosssection1}
when the annihilation happens around the resonance pole,
$s \sim m^2_\phi$ such that
$\beta_{i,f} \sim \bar{\beta}_{i,f}$.
Then most of the $\beta$ factors would cancel out and the prefactor
$\beta_i \beta^3_f / \bar \beta^3_i \bar \beta^3_f$
here approaches the $1/\bar \beta^2_i$ factor in
\geqn{eq:crosssection1}. This is a good enough
approximation, especially for a narrow resonance peak
$\gamma_\phi \lesssim 10^{-5}$ shown in
\gfig{fig:RelicDensiy}.

The scalar coupling $\lambda$ with the SM Higgs
boson should maintain thermal equilibrium for $\phi$ in
the early Universe. In other words, 
the decay rate of $\phi$ to the SM fermions $\Gamma_{\phi \rightarrow f \bar f}$
should be larger than the Hubble rate $H \propto T_f^2/M_{\rm pl}$.
Suppose $\phi$ decouples around 
the freeze-out temperature $T_f \sim m_\phi / 25$,
thermal equilibrium requires $\lambda \gtrsim (10^{-8} \sim 10^{-7})$
with the concrete value
depending on the mediator mass $m_\phi$. 
Further through the first Yukawa term in \geqn{Interaction} and the resultant
$\phi \rightarrow N_\chi + N_\chi$ decay, $N_\chi$
can also get in thermal equilibrium. 
The parameter space
of this RHN DM model to give the correct 
relic density can be directly read off from \gfig{fig:RelicDensiy}. 
Across the whole parameter space, the 
coupling strength remains perturbative.

{\bf Reactivation around SMBH} --
In the current Universe, the DM 
particles have already become
non-relativistic,
which means almost no DM particles can gain enough 
kinetic energy to reach the  Breit-Wigner resonance pole.
Thus, the DM annihilation cross section is 
highly suppressed to leave almost no 
signal in indirect detection.
However, a strong gravitational source, such as
an SMBH, can accelerate the DM particles
to reactivate their annihilation
\cite{Cheng:2022esn}. For illustration, the red
curve in \gfig{fig:BWCurve} shows the velocity
distribution at the radius $r$ where the DM
annihilation rate per unit radius, 
$4 \pi r^2 \rho_\chi^2(r) \langle \sigma v_{\rm rel} \rangle 
/ m_{\chi}^2$, is maximized.
Here, $\rho_\chi (r)$ is the DM density around the SMBH
\cite{Fields:2014pia,Alvarez:2020fyo}. 
We can see that the DM velocity distribution near SMBH
(red dashed)
also has a big overlap with the Breit-Wigner resonance peak
(black solid). 
Therefore, the DM will gradually annihilate into 
the SM fermions around SMBH and then subsequently 
decays into gamma rays as observable signal.

The observed gamma ray flux
$d \Phi_{\gamma}/d E_{\gamma}$
is an integration 
of the produced differential gamma ray flux
$dF_\gamma / dE_\gamma$
over the radius $r$,
\begin{equation}
  \frac{d \Phi_{\gamma}}{d E_{\gamma}}
=
  \frac{1}{4 \pi D^2} \frac{1}{2 
  m_\chi^{2}}
  \int_{r_c}^{r_b} 4 \pi r^{2} d r  
  \rho^{2}_\chi (r) \frac{dF_{\gamma}}{dE_{\gamma}}(r),
\label{eq:gamma_flux}
\end{equation}
where $D = 8.5$\,kpc is the distance between the Milky Way galaxy center
and our solar system.
For the innermost capture region,
$r < r_c \equiv 4GM $ where $G$ is the Newtonian constant \cite{Sadeghian:2013laa,Shapiro:2014oha}, all particles are
attracted to fall into the SMBH so that there is no DM.
The SMBH at the center of our galaxy has a mass
$M = 4 \times 10^6 M_{\odot}$ almost four millions
of the solar mass $M_\odot$.
When the gravitational
influence of SMBH no longer dominates for
$r > r_b \equiv 0.2 G M/v_0^2
$ \cite{Gondolo:1999ef} where $v_0 = 105\,$km/s \cite{Gultekin:2009qn} is the DM
velocity dispersion outside the spike, the DM halo simply
follows the NFW profile \cite{Navarro:1995iw}.
In between, a DM spike forms.
In principle, we should include all the contributions along the line of sight by integrating all the way from $r_c$ to $D$. However, 
the annihilation outside
the spike is negligible
due to lack of the Breit-Wigner
enhancement. So we take the integration upper limit in \geqn{eq:gamma_flux} as
the outer boundary $r_b$ of spike.

The full DM profile
around a SMBH is,
\begin{equation}
\hspace{-4mm}
  \rho_\chi(r)
=
\begin{cases}
  0,
& r < r_c, \text { (Capture Region),}
\\
  \frac{\rho_{\mathrm{sp}}(r) \rho_{\mathrm{in}}(t, r)}{\rho_{\mathrm{sp}}(r)+\rho_{\mathrm{in}}(t, r)},
& r_c \leq r < r_b,
  \text { (Spike), }
\\
  \rho_b (r_b/r)^{\gamma_c},
&
 r_b < r < D,
  \text { (Halo). }
\label{CDM_Profile}
\end{cases}
\end{equation}
The handy spike profile 
$\rho_{\rm sp} (r) \equiv \rho_b (r_b/ r)^{\gamma_{\rm sp}}$  scales from the density
$\rho_b \equiv 0.3\,$GeV/cm$^3 \times (D/r_b)^{\gamma_c}$
with $\gamma_{c} = 1$ 
according to the NFW profile
\cite{Navarro:1995iw}
at the spike boundary. The radius scaling power index $\gamma_{\rm sp}$ strongly 
depends on the formation history of the SMBH. 
For example,
the adiabatic growth of the central 
SMBH gives the steepest spikes profile, $\gamma_{\rm sp} = (9 - 2 \gamma_c)/(4-\gamma_c) \sim 2.3$ \cite{Gondolo:1999ef} with only $1$\% variation.
To avoid too aggressive predictions for the DM density profile and the subsequent $\gamma-$ray signal, we take $\gamma_{\rm sp} = 1.8$  as a conservative benchmark value \cite{Gnedin:2003rj,Merritt:2003qk,Bertone:2005hw,Merritt:2006mt,Shelton:2015aqa,Johnson:2019hsm,Chiang:2019zjj} for the more realistic non-adiabatic evolution.
For the inner profile, the density becomes
$\rho_{\rm in} (r) \equiv \rho_{\rm ann} (r/r_{\rm in})^{-\gamma_{\rm in}}$ with $\gamma_{\rm in} = 1/2$. The radius
$r_{\rm in}$ determined by
$\rho_{\rm sp}(r_{\rm in}) = \rho_{\rm ann}$
where the plateau density $\rho_{\rm ann}$ is 
$\rho_{\rm ann} \equiv m_\chi / 
\braket{\sigma v_{\rm rel}} \tau$ \cite{Vasiliev:2007vh} with the galaxy lifetime $\tau$.
This is because only a certain relative velocity can hit the resonance pole with a very 
tiny relative mediator decay width $\gamma_\phi \lesssim 10^{-5}$ and consequently the annihilation
cross section is effectively $s$-wave with no actual velocity dependence.

\begin{figure}[!t]
\centering
 \includegraphics[width=0.486
 \textwidth]{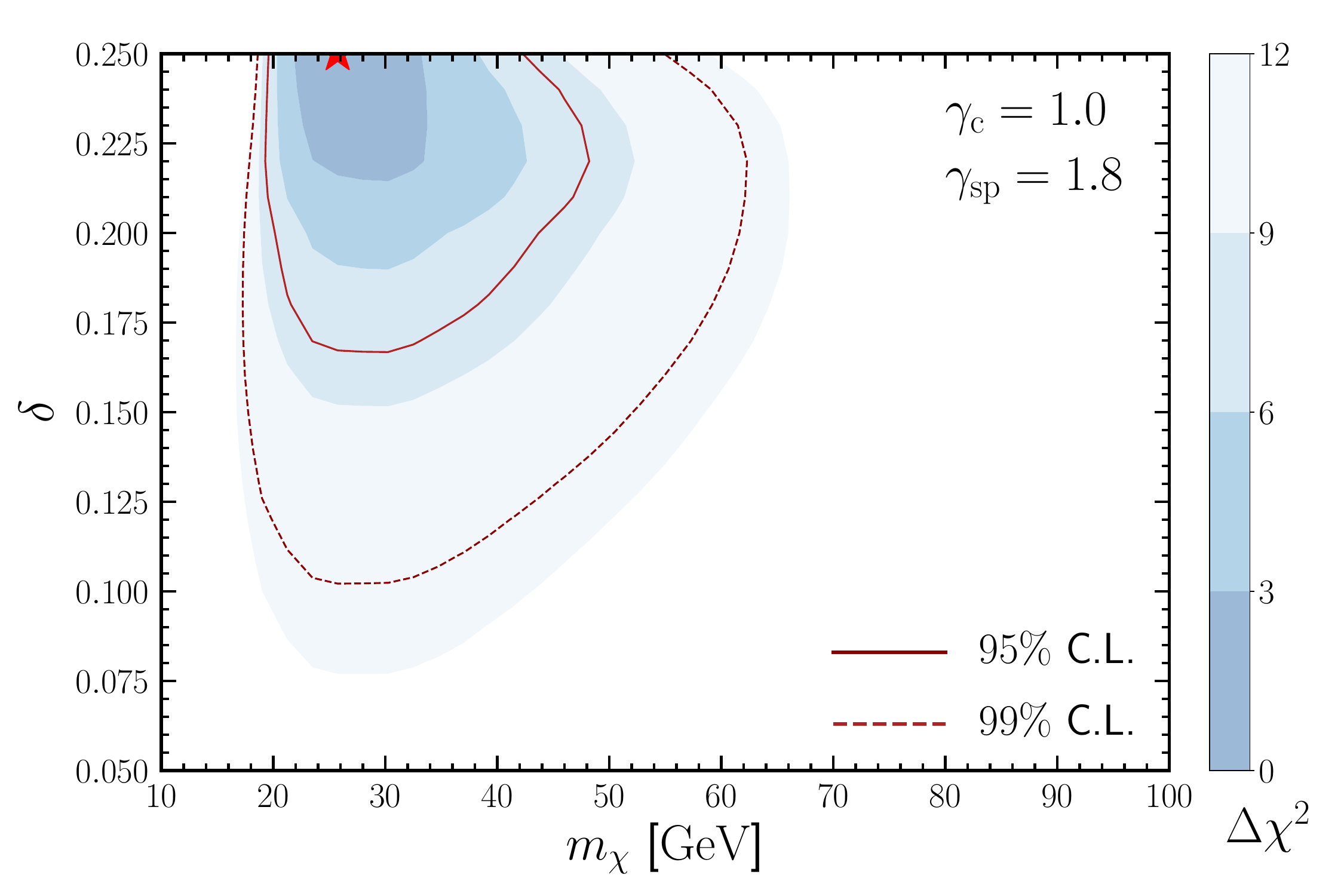}
\caption{
The 95\% (red solid) and 99\%\,C.L. (red dashed) sensitivity contours for different DM masses $m_\chi$ and mass differences $\delta$ obtained by fitting the Fermi-LAT data. 
We take $B_i B_f = 0.01$, $\gamma_c = 1$, and $\gamma_{\rm sp} = 1.8$ as an illustration. 
The best-fit point (red star) is at $m_\chi = 26\,$GeV and $\delta = 0.25$
with $\chi^2_{\rm min} = 32$.
The shaded regions in different colors represent different 
$\Delta \chi^2$ ranges.
}
\label{fig:BHConstraint}
\end{figure}

The differential gamma ray flux
$dF_\gamma / dE_\gamma$ in \geqn{eq:gamma_flux} from the DM annihilation and subsequent decay is, 
\begin{eqnarray}
\hspace{-2mm}
  \frac{dF_{\gamma}}{dE_{\gamma}}
= 
\hspace{-1mm}
\int_0^1 
\hspace{-1mm}
d V_{r} d V_{c}  \mathcal{P}_{\mathrm{r}}
 \left(V_{r}, V_c\right)   (\sigma v_{\rm rel})_{\rm cm}
 \frac{dN_{\gamma}}{dE_{\gamma}}(V_{r},V_c).
\label{eq:relativistic_flux_complex}
\end{eqnarray}
Finally expressed in terms of the DM energy
$E_i$ ($i = 1,2$) and momentum $p_i$,
the joint probability 
distribution 
$\mathcal{P}_{\mathrm{r}}$ for relativistic thermalized particles \cite{Cannoni:2013bza,Cannoni:2015wba},
\begin{equation}
\hspace{-2mm}
\mathcal{P}_{\mathrm{r}}
  \left(V_{r}, V_c\right)
  =
  \frac{x^2 \gamma_{r}^2 
  (\gamma_{r}^2-1) (1 + \gamma_r) V^2_c}
  {2 K_2^2(x) (1-V_c^2)^{2}}
  e^{-x \sqrt{\frac{2  + 2  \gamma_r}{ 1-V_c^2}}
  },
  \label{eq:PrVrVc}
\end{equation}
can be more conveniently expressed as
a function of
their relative velocity $V_r \equiv \sqrt{\gamma^2_r - 1} / \gamma_r$
($\gamma_r \equiv (p_1 \cdot p_2) / m^2_\chi$)
and the center-of-mass velocity 
$V_c \equiv \sqrt{1 - s / (E_1 + E_2)^2}$.
The temperature parameter 
$x(r) \equiv m_\chi/T_\chi(r)$ now has radius $r$ dependence through the 
DM temperature $T_\chi (r) = \frac12 m_\chi 
v_d^2 (r)$ where $v_d (r)$ is 
the DM velocity dispersion.
Its radius dependence is approximately 
$v_d (r) \propto v_0 (r/r_b)^{-1/2}$ which 
can be deduced from the Virial theorem $v_d^2 (r) \propto G M/r$ \cite{Shelton:2015aqa,Chiang:2019zjj}. 
For $m_\phi > 10\,$GeV, the mediator $\phi$ produced from the DM annihilation first mainly
decays into $\phi \rightarrow b \bar b$. The following processes, such as quark and lepton radiation, and $\pi_0$ decay, can generate a photon flux $dN_\gamma/dE_\gamma$ \cite{Cirelli:2010xx}.
Note that the cross section
$ (\sigma v_{\rm rel})_{\rm cm} $ is defined 
in the center-of-mass frame of DM 
collision. 
We use the {\tt PPPC4DMID} package
\cite{Cirelli:2010xx} to generate the photon spectra 
and further boost it to the lab frame
\cite{Elor:2015bho,Elor:2015tva}.

We use 15
years of Fermi-LAT data \cite{DataAcess} from August 4, 2008 to May 26, 2023 and analyze the $10^\circ \times 10^\circ$ square region centered around Sgr\,$A^{*}$ with pixel size $0.08^\circ \times 0.08^\circ $ \cite{Analysis}.  Among all the $\gamma$-ray sources inside this region, we select 
the point source 4FGL\,J1745.6-2859 that is the brightest and closest one to Sgr\,$A^*$ 
using the official package Fermitools \cite{Fermitools} or Fermipy \cite{Wood:2017yyb}
after removing the diffusion background using 
the \texttt{Pass 8 SOURCE event class} \cite{Bruel:2018lac}.
This point source from the Fourth 
catalog of Fermi-LAT sources (4FGL) \cite{Fermi-LAT:2019yla,Ballet:2020hze}
is considered as the manifestation of Sgr\,$A^*$ \cite{Cafardo:2021pqs}.
In our analysis, those events from
100\,MeV to 100\,GeV are binned into 12 logarithmically
spaced energy bins.

To describe the $\gamma$-ray background spectrum from
the point source 4FGL\,J1745.6-2859,
we take a univeral background model with log-parabola function
\begin{equation}
  \frac{d N}{d E}
=
  N_0 \left( \frac E {E_{0}} \right)^{-\alpha-\beta \log (E / E_{0})},
  \label{eq:BCKmodel}
\end{equation}
where $N_0$ is the normalization factor, $E_0$ is a scale parameter,
$\alpha$ gives the spectral slope at $E_0$, and $\beta$ measures the curvature of the spectrum \cite{Massaro:2003sx,Cafardo:2021pqs}. The value of $E_0$ does not vary much and is fixed \cite{Massaro:2003sx} to 6499\,MeV, while the other three can freely adjust.

For each parameter space point generating the right DM relic in \gfig{fig:RelicDensiy}, the predicted $\gamma$-ray flux around the SMBH is uniquely determined. Combining the DM annihilation signal with the background model in \geqn{eq:BCKmodel} and performing $\chi^2$ minimization, we find the best-fit point at $m_\chi = 26\,$GeV and $\delta = 0.25$ with the corresponding $\chi^2_{\rm min} = 32$ shown as red star in \gfig{fig:BHConstraint}.
According to the 95\%\,C.L. (red solid) and
99\%\,C.L. (red dashed) limits,
the allowed parameter space is within 
$m_\chi \subset (20, 60)\,$GeV and $\delta \subset (0.1, 0.25)$.
For smaller mass difference $\delta$, the annihilation around SMBH becomes
more significant. Being combined with the background,
the DM annihilation signal exceeds the Fermi-LAT data.
The best-fit DM mass $25\,$GeV generates a photon
flux with peak energy $m_\chi/10$
\cite{Cirelli:2010xx} around the excess observed
in the Fermi-LAT data.

{\bf Conclusions and Discussions}
--
We propose for DM annihilation a 
new Breit-Wigner enhancement scenario
with the mediator mass larger 
than two times of the DM mass. 
While the DM annihilation
is highly suppressed to 
evade the CMB constraint when the 
temperature cools down, the DM
freeze-out in the early Universe and its 
reactivation around an SMBH are enhanced
by the $s$-channel resonance with large 
enough kinetic energy.
For illustration, we construct a 
UV complete RHN DM model for 
the DM mass range from $\mathcal{O}(0.1)$\,GeV
to $\mathcal{O}(100)$\,GeV.


\section*{Acknowledgements}

Yu Cheng and Jie Sheng 
would like to thank Prof. Shigeki Matsumuto for useful discussions and hospitality during their
stay at Kavli IPMU where this paper was partially completed.
SFG is supported by the National Natural Science
Foundation of China (12375101, 12425506, 12090060, and 12090064) and the SJTU Double First
Class start-up fund (WF220442604).
T. T. Y. is supported by the China Grant for
Talent Scientific Start-Up Project and by Natural Science
Foundation of China (NSFC) under grant No.\,12175134,
JSPS Grant-in-Aid for Scientific Research
Grants No.\,19H05810, 
and World Premier International Research Center
Initiative (WPI Initiative), MEXT, Japan.
Both SFG and T. T. Y.
are affiliated members of Kavli IPMU, University of Tokyo.

\section*{Appendix: Intuitive Definition of Dark Matter Annihilation Quantities}

In the Boltzmann equation, 
the change rate $\dot n$ of the
particle number density $n$ is 
typically derived from the scattering 
cross section. However, its definition in 
an arbitrary frame with Moller velocity \cite{Gondolo:1990dk} 
or a Lorentz invariant $V_r$ \cite{Cannoni:2013bza,Cannoni:2015wba} is 
not intuitive. A more natural way is 
using the geometrical picture of cross section
in the center-of-mass frame with head-on 
collision \cite{Peskin:1995ev}. The event rate is then
a product of particle number densities $d n_i$, 
cross section $\sigma$,
and the relative velocity $v_{\rm rel}$ therein,
\begin{equation}
   d R =  (\sigma v_{\rm rel})_{\rm cm} d n_1^{\rm cm} d n_2^{\rm cm}
      =  \frac{E_1^{\rm cm}E_2^{\rm cm}}{E_1 E_2}
        (\sigma v_{\rm rel})_{\rm cm} d n_1 d n_2.
    \label{eq:CollsionRate}
\end{equation}
Note that $d n_i \equiv f_i({\bm p}_i) d^3 {\bm p}_i$
is the number density element around momentum
${\bm p}_i$ such that all particles
enclosed in the same $d n_i$ has exactly the same
the kinematics.
The transition from the particle number densities
$d n_i^{\rm cm}$ in the center-of-mass frame to the commonly used ones $d n_i$
in the cosmic frame is simply a Lorentz boost
factor $\gamma_i \equiv E^{\rm cm}_i/E_i$.
It is not just that this conceptual line 
is very intuitive, 
the calculation of cross section as well as 
the corresponding differential spectrum in 
the center-of-mass frame is much simpler. 
For example, a $2 \rightarrow 2$ scattering 
has mono-energetic final-state particles and the differential spectrum in the cosmic frame
can then be obtained with a Lorentz boost.
For isotropic scattering in the center-of-mass
frame, such Lorentz boost would render a box-
shaped spectrum from 
a mono-energetic final-state particle.

For the head-on $2 \rightarrow 2$ collision,
the cross section times relative velocity is, 
\begin{align}
\hspace{-2mm}
  (\sigma v_{\rm rel})_{\rm cm}
\equiv&
\frac{1}{4 E_1^{\rm cm} E_2^{\rm cm} g_1 g_2}
\sum_{\text {spins }} \int
\frac{d^3 {\bm p}_3}{(2 \pi)^3 2 E_3} \frac{d^3 {\bm p}_4}{(2 \pi)^3 2 E_4}\nonumber\\
&(2 \pi)^4 \delta^4\left(p_1+p_2-p_3-p_4\right) 
\left|\mathcal{M}_{12 \rightarrow 34}\right|^2,
\label{eq:sigmavCM}
\end{align}
where $g_1$ and $g_2$ come from the spin average
of initial particles.
We can see that $E_1^{\rm cm} E_2^{\rm cm}$
cancels with those in \geqn{eq:CollsionRate} that originate
from the Lorentz boost factors $\gamma_i$.
Then it is possible to calculate the event rate
in the lab frame, $(\sigma v_{\rm rel})_{\rm lab} \equiv
(\sigma v_{\rm rel})_{\rm cm} \times (E_1^{\rm cm} E_2^{\rm cm}
/ E_1 E_2)$,
which is convenient in the sense that there is
no need to involve the individual cross section
or relative velocity but their product. The explicit
form of $(\sigma v_{\rm rel})_{\rm lab}$ resembles
\geqn{eq:sigmavCM} with $E^{\rm cm}_i$ replaced by
$E_i$.
Whether using $(\sigma v_{\rm rel})_{\rm cm}$ or
$(\sigma v_{\rm rel})_{\rm lab}$ for concrete calculation
is up to convenience. It is conceptual intuitive
and computationally convenient enough.

The formalism with Moller velocity $v_{\rm mol} \equiv F/ E_1 E_2$ \cite{Gondolo:1990dk}
can be reproduced by inserting a
flux factor $F \equiv \sqrt{\left(p_1 \cdot p_2\right)^2-m_1^2 m_2^2}$. 
Correspondingly, the cross section reduces to 
$\sigma_{\rm Moller} \equiv \sigma_{\rm cm} \times (E_1 E_2/ F)$
such that the cross section
times the relative velocity is the same
$\sigma_{\rm Moller} v_{\rm mol} = \sigma_{\rm cm} v_{\rm cm}$.
For its variant, the Moller velocity is replaced by $v_{\rm mol} = (p_1 \cdot p_2 / E_1 E_2) \times V_r$ with a Lorentz invariant $V_r \equiv F / (p_1 \cdot p_2 )$ 
and the corresponding cross section is 
$\sigma_{\rm cm} \times (p_1 \cdot p_2 / F)$
\cite{Cannoni:2013bza,Cannoni:2015wba}.
In those definitions, neither the 
velocity has the meaning
of physical relative velocity nor the 
cross section has geometrical picture.
Except the only advantage of being 
Lorentz invariant for $\sigma_{\rm Moller}$ and $V_r$,
such 
definitions are actually quite abstract comparing with
the intuitive definitions in the center-of-mass frame.

\providecommand{\href}[2]{#2}\begingroup\raggedright\endgroup

\vspace{15mm}
\end{document}